Preprint -- Submitted to *Magnetic Resonance in Medicine*

# A Digital Phantom for MR Spectroscopy Data Simulation

**D.M.J. van de Sande[1], A.T. Gudmundson[2,3,4], S. Murali-Manohar[2,3], C.W. Davies-Jenkins[2,3], D. Simicic[2,3], G. Simegn[2,3], İ. Özdemir[2,3], S. Amirrajab[1], J.P. Merkofer[5], H.J. Zöllner[2,3], G. Oeltzschner[2,3], R.A.E. Edden[2,3]**

[1](Department of Biomedical Engineering, Eindhoven University of Technology, Eindhoven, The Netherlands)
[2](Russell H. Morgan Department of Radiology and Radiological Science, The Johns Hopkins University School of Medicine, Baltimore, Maryland, USA)
[3](F. M. Kirby Research Center for Functional Brain Imaging, Kennedy Krieger Institute, Baltimore, Maryland, USA)
[4](The Malone Center for Engineering in Healthcare, Johns Hopkins University School of Engineering, Baltimore, Maryland, USA)
[5](Department of Electrical Engineering, Eindhoven University of Technology, Eindhoven, The Netherlands)

## Abstract

**Purpose**
Simulated data is increasingly valued by researchers for validating MRS processing and analysis algorithms. However, there is no consensus on the optimal approaches for simulation models and parameters. This study introduces a novel MRS digital brain phantom framework, providing a comprehensive and modular foundation for MRS data simulation.

**Methods**
The framework generates a digital brain phantom by combining anatomical and tissue label information with metabolite data from the literature. This phantom contains all necessary information for simulating spectral data. The MRS phantom is combined with a signal-based model to demonstrate its functionality and usability in generating various spectral datasets. Outputs can be saved in the NIfTI-MRS format, enabling their use in downstream applications. To evaluate the realism of the simulated spectra, a comparison was performed against in-vivo MRS data acquired under similar conditions.

**Results**
The phantom was implemented using two anatomical templates at different resolutions and tested across a range of user-defined simulation parameters. Simulated spectra exhibited realistic signal characteristics and structural variability. When compared to in-vivo data, the simulated spectra closely matched in terms of spectral shape, signal-to-noise ratio, and metabolite quantification. The simulations also captured key variability features and provided additional diversity not present in the in-vivo dataset, supporting use in robustness testing and data augmentation.

**Conclusion**
This novel digital phantom provides a flexible and extensible platform for MRS data simulation. Its modular architecture, user-friendly GUI, and open-source implementation support reproducible research, algorithm development, and validation in the MRS community.

# 1 Introduction

Magnetic resonance spectroscopy (MRS) is a non-invasive technique for measuring neurochemical concentrations in the human brain, offering valuable insights into biochemical changes linked to neurological conditions[1,2]. Despite their promise, the clinical implementation of MRS has been hindered by several challenges, including time-consuming acquisitions, inherently low signal-to-noise ratios (SNRs), and the specialized expertise required for data processing and analysis[3–5].

To address these challenges, researchers are developing methodologies and toolboxes to streamline and standardize the processing and analysis of MRS data[6–9], including machine learning approaches[10–12]. Progress in these areas requires sufficient data for validation, yet access to high-quality MRS datasets is limited due to the scarcity of open databases, high acquisition costs, and the absence of ground-truth concentrations in in-vivo data.

Consequently, simulating MRS datasets is increasingly recognized as a valuable practice among researchers. Synthetic data generation not only allows for the creation of a vast number of spectra but also enables control over the ground-truth values associated with these spectra. Despite the widespread use of spectral simulation, there is no consensus on the optimal approaches for simulation models and parameters. The choices of signal model, parameter ranges, and validation of data realism are often intricate and diverse.

Within the field of MRI, numerous digital phantoms have been developed to simulate MR images[13–16]. These phantoms effectively combine anatomical information and tissue properties with physics-based models to produce realistic MR images. Such simulation tools have significantly advanced research in MRI image generation and analysis. To our knowledge, no digital phantom exists that is specifically designed to generate MRS data by integrating anatomical information with current knowledge of brain metabolites. Therefore, the purpose of this work is to develop a novel simulation framework for generating MRS data using a digital brain phantom. This framework focuses on integrating anatomical brain information with established literature regarding brain metabolite concentrations and relaxation times. The MRS digital brain phantom framework provides a comprehensive and modular tool for data generation, with a user-friendly GUI that broadens accessibility and enables use across varying levels of expertise. The framework is evaluated by comparing simulated spectra to in-vivo data and assessing its flexibility in generating both realistic and augmented datasets. The full implementation is openly available in Python to support future development and reproducible research in the MRS community (https://github.com/dennisvds/MRS-Digital-Phantom).

# 2 Materials & Methods

The overall outline of our proposed MRS digital brain phantom framework is shown in Figure 1. This structure is divided into three stages: skeleton, MRS phantom, and simulation. All computational tasks were performed using a MacBook Pro with M2 Pro chip (10 core CPU/16 core GPU) and equipped with 16GB of RAM.

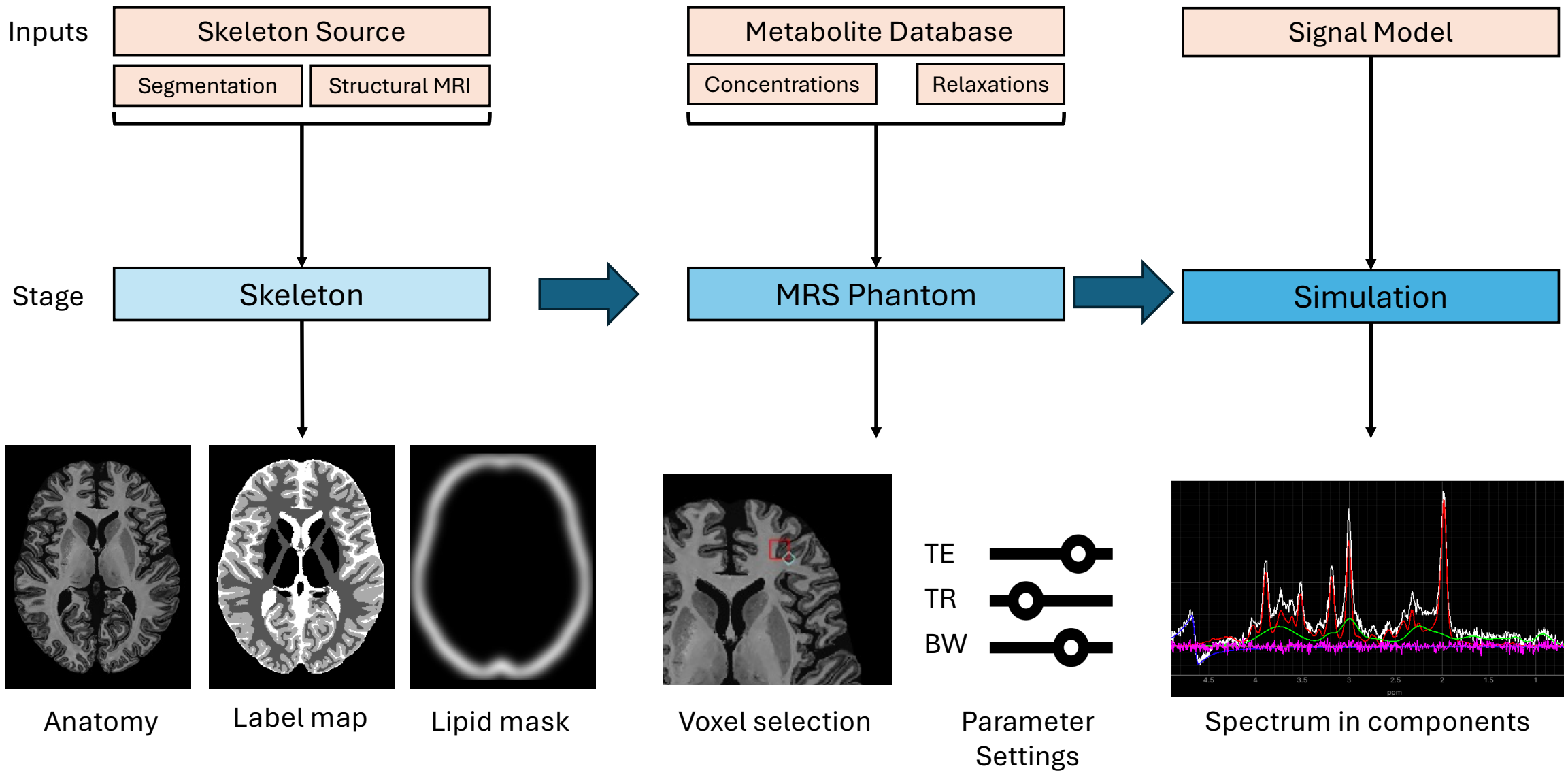

*Figure 1: Structure of the 3D MRS digital brain phantom framework. The framework is divided into three stages: Skeleton, MRS Phantom, and Simulation. Each stage allows for user-defined inputs, making the framework highly modular and customizable for various applications. This example uses the 'BigBrain-MR' skeleton source.*

## 2.1 Skeleton

The anatomical foundation of the proposed framework is a three-dimensional brain phantom, referred to as the skeleton. This skeleton encodes anatomical structure and tissue labels for the MRS phantom. It also serves as the basis for deriving a lipid mask, which models extracranial lipid contamination. Any anatomical phantom that includes a structural MRI and corresponding tissue segmentation can serve as input for the MRS phantom. In this study, two sources of brain anatomical data are implemented: the MRiLab phantom and BigBrain-MR.

The MRiLab phantom is sourced from a MATLAB-based numerical MRI simulation package called MRiLab[13]. It includes a three-dimensional tissue label map with 1 mm resolution and a corresponding proton density (PD) map. The label map contains classifications for white matter (WM), gray matter (GM), cerebrospinal fluid (CSF), skull, and fat. The lipid mask $M_L(x, y, z)$ is generated by extracting the binary fat label map and applying a Gaussian filter. The standard deviation $\sigma$ of this Gaussian kernel, defined in millimeters, is converted into voxel units using the known voxel spacing. The smoothed result is normalized to yield a soft spatial map indicating proximity to lipid-rich regions. During single-voxel spectroscopy (SVS) voxel selection, this spatial map is used to guide placement, ensuring realistic lipid contamination in voxels located near the skull.

BigBrain-MR is a high-resolution three-dimensional brain phantom, constructed by integrating a 7T low-resolution MRI with a histological dataset to create an *in-vivo*-like map of tissue properties[16]. Although the original resolution reaches up to 100 μm, a downsampled version at 400 μm resolution is employed to reduce computational requirements. Of the 20 tissue labels included in the BigBrain-MR phantom, only those corresponding to WM, GM, and CSF are utilized. As the phantom is skull-stripped, it lacks explicit information regarding skull and fat tissue. To estimate the lipid mask $M_L(x, y, z)$, it is assumed that lipid-rich regions lie along the outer brain boundary. A total brain mask is eroded and subtracted from the original brain mask, resulting in an estimated lipid mask surrounding the brain. This mask is subsequently smoothed using a Gaussian kernel with standard deviation $\sigma$ and normalized similarly to the lipid mask for the MRiLab skeleton.

All phantom maps derived from BigBrain-MR and MRiLab are saved in the NIfTI file format[17]. For integration into the MRS phantom workflow, these maps are loaded using the TorchIO library[18].

## 2.2 MRS Phantom

### 2.2.1 Metabolite Database

In the second stage, metabolite information is combined with the anatomical skeleton. This study uses metabolite concentrations and $T_2$ values for WM and GM are extracted from an open-source database of a previously published meta-analysis[19], which summarizes nearly 500 studies reporting metabolite relaxation times and concentrations in healthy and pathological brains.

The database is processed following Figure 2. All studies are filtered on healthy and control patients and are only included when information about GM and WM fractions is present. Since the phantom uses binary tissue labels, metabolite concentrations were assigned to GM or WM based on tissue fractions ≥ 0.6. Studies were further filtered by age (18–60 years), and metabolite nomenclature was unified. The metabolite concentrations for tCr, tNAA, tCho, and Glx are split into their individual components based on known relations found in literature[20,21] , and the concentration data in units of mM and IU are combined similarly as in the meta-analysis. The $T_2$ database is filtered on studies that use 3T scanners. Finally, metabolite concentrations and $T_2$ values are calculated using a random effects model[22], with weights determined by the inverse square of the reported standard deviations. If only one study is available for a specific tissue-metabolite combination, the values reported in this study are used.

When all metabolite concentrations and $T_2$ relaxation times are calculated, the results are saved in a metabolite dataframe format that is integrated into the MRS phantom. It ensures compatibility with the MRS phantom, but also allows the user to choose their own metabolite concentrations and relaxation times within this specified format. Since the meta-analysis did not contain any information about CSF labelled voxels, metabolite information about these voxels is manually added in the metabolite dataframe based on other literature values[23]. The framework also requires background labels, which represent empty space. All metabolite values for these labels are automatically set to zero. An example of this metabolite dataframe format is shown in Supporting Information Table S1.

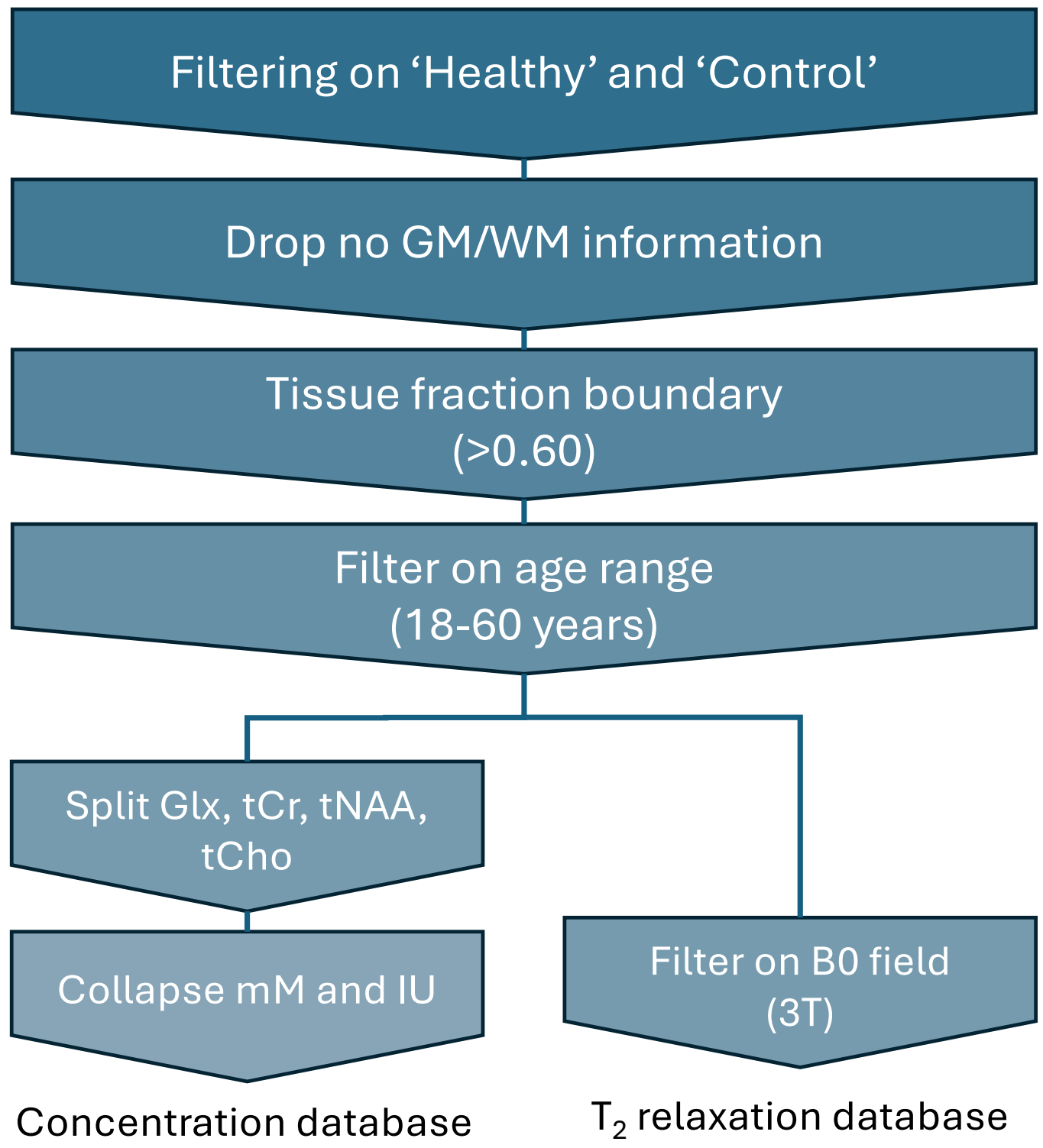

*Figure 2: Filtering flowchart of the metabolite database entries used. These steps are performed to select the appropriate subset of entries collected in the previous meta-analysis.*

### 2.2.2 Phantom Class

The MRS phantom is the core structural component of the framework, combining anatomical skeleton information with metabolite data. It is implemented as a Python class, '*DigitalPhantom*', which contains the skeleton, metabolite dataframe, and all relevant simulation parameters. These parameters can be specified via a configuration file or interactively through a graphical user interface (GUI). Table 1 summarizes the available configuration options.

The volume of interest (VOI) is defined by voxel indices specifying minimum and maximum boundaries along the x-, y-, and z-axes. Selection of the VOI is aligned with the underlying voxel grid of the phantom, ensuring that the VOI boundaries always coincide with voxel edges. VOI selection aligns with the phantom’s voxel grid, ensuring boundaries coincide with voxel edges. This prevents partial voxels at the VOI edges and maintains internal consistency. VOI selection can be configured manually in the configuration file or interactively via the GUI’s placement tool.

In this framework, “VOI” is explicitly used to avoid ambiguity. Unlike traditional SVS, where a voxel denotes the measurement region, the phantom operates at a finer spatial resolution, so a single VOI contains multiple phantom voxels.

*Table 1: Overview of all parameters relevant for the Digital MRS Phantom and its simulation model. These parameters can all be defined by the user.*

| Parameter | Type | Description |
|---|---|---|
| | **General Inputs** | |
| skeleton | string | Specifies the anatomical phantom. Options: "BigBrainMR", "MRiLab". |
| path2metabs | string | Path to the metabolite dataframe. |
| basis_set_dir | string | Directory containing the basis sets. |
| path2basis | string | Path to the basis set. Can be used to load a specific basis set. |
| metabs | list of strings | List of metabolites to include. Leave empty to include all available metabolites. |
| | **Basis Set Settings** | |
| vendor | string | MRI system vendor for basis set generation. Options: "Universal_Philips" or "Universal_Siemens”. |
| localization | string | Localization technique. Options: "PRESS", "STEAM", "sLASER". |
| TE | float | Echo time (ms). |
| | **Simulation Parameters** | |
| spectral_points | integer | Number of points in the simulated spectrum. |
| TR | float | Repetition time (ms). |
| bandwidth | float | Spectral bandwidth (Hz). |
| noise_level | float | Standard deviation of sub-voxel Gaussian noise. |
| mm_level | float | Amplitude scaling factor for macromolecule signal |
| mm_json_file | string | Path to configuration file for macromolecule simulation. |
| lipid_amp_factor | float | Amplitude scaling factor for lipid signals. |
| lipid_sigma | float | Standard deviation (in mm) of the Gaussian smoothing kernel applied to the lipid mask. |
| lipid_phase_min, lipid_phase_max | float | Range of random phase shifts (in degrees) applied per lipid component. |
| lipid_lw_min, lipid_lw_max | float | Linewidth range (in Hz) for exponential decay per lipid component. |
| water_amp_factor | float | Amplitude scaling factor for water signal. |
| water_phase_min, water_phase_max | float | Phase range for water signal. |
| water_damping_min, water_damping_max | float | Damping range for water signal. |
| shim_amplitude_hz | float | Amplitude scaling factor for shim imperfection map (in Hz). |
| shim_corr_length | float | Correlation length (mm) for simulated shim imperfections. |
| shim_boundary_amp_factor | float | Scaling factor for boundary-related shim contributions. |
| shim_boundary_smoothing | float | Smoothing factor applied to boundary-related shim contributions. |
| | **VOI Definition** | |
| coords | list of integers | Minimum and maximum voxel indices per dimension : [x_min, x_max, y_min, y_max, z_min, z_max]. |
| size | list of integers | Size of the VOI in voxel units [x,y,z]. |
| size_mm | list of floats | Size of the VOI in millimeters [x_mm, y_mm, z_mm]. |

## 2.3 Signal Model

The final element that completes the MRS phantom structure is the signal model. The signal model takes all information stored in the MRS phantom and uses it to generate SVS MRS data. The signal model is in itself a modular design and can be written down as

$$S(\nu) = S_{\mathrm{met}}(\nu) + S_{\mathrm{mm}}(\nu) + S_{\mathrm{lipids}}(\nu) + S_{\mathrm{H2O}}(\nu) + N(\nu) \quad (1)$$

with $S(\nu)$ the total spectrum in the frequency domain and $S_{met}(\nu)$, $S_{mm}(\nu)$, $S_{lipids}(\nu)$, $S_{H2O}(\nu)$, and $N(\nu)$ the corresponding metabolite, macromolecule, lipid, water and noise components. The rest of this section will discuss each component in more detail.

### 2.3.1 Metabolite Simulation

To simulate metabolite spectra in the frequency domain, a set of basis functions $f_k(t)$ is used for each metabolite $k$. For each tissue label $l$, the corresponding time-domain free induction decay (FID) signal is modeled as:

$$\mathrm{FID}_{\mathrm{met},kl}(t) = f_k(t)\exp(-\gamma_{kl}t) \quad (2)$$

Here, $\gamma_{k\ell}$ is the Lorentzian broadening constant, derived from the $T_2$-values specified the metabolite dataframe via the relation $\gamma_{k\ell} = \frac{1}{\pi T_2}$. The resulting signal is then Fourier-transformed, scaled by the tissue-specific metabolite concentration $c_{k\ell}$, and summed over all metabolites:

$$S_{\mathrm{met,l}}(\nu) = c_{kl} \cdot \mathcal{F}\left[\mathrm{FID}_{\mathrm{met},kl}(t)\right]. \quad (3)$$

The values for $c_{k\ell}$ are obtained by random sampling from a Gaussian distribution defined by the mean and standard deviation provided in the metabolite dataframe. for This tissue-specific spectrum is then weighted using a one-hot encoded tissue label map $L_l(x, y, z)$ of the selected VOI, and averaged over the full VOI:

$$S_{\mathrm{met}}(\nu) = \left\langle \sum_{l} L_l(x, y, z) \cdot S_{\mathrm{met},l}(\nu) \right\rangle_{x,y,z} \tag{4}$$

where $\langle \cdot \rangle_{x,y,z}$ denotes the spatial average over the entire VOI. The basis functions $f_k(t)$ are provided in the '.basis' file format and are loaded according to the selected simulation settings (see Table 1), such as localization method and echo time. Users may supply a custom basis set, or alternatively, the framework can generate one automatically using MRSCloud[24], based on the selected vendor, localization, and TE. This automated generation currently supports PRESS localization using universal pulse sequences. If a previously generated basis set matching the selected parameters is available, it is reused to avoid redundant calculations.

### 2.3.2 Macromolecule Simulation

The macromolecular (MM) background is modeled using the methodology described by Wright et al.[25]. This approach generates a sequence-specific, relaxation-corrected MM spectrum that serves as a basis MM spectrum in the overall signal model. Each MM component is represented as a Voigt profile, incorporating both Lorentzian and Gaussian linewidth components, and modulated by tissue type and sequence parameters.

To simulate the MM background at 3T, parameter values were sourced from previously published experimental studies. Measured MM linewidths, water linewidths (used to estimate shim-related broadening), and $T_2$ relaxation times for ten MM resonances in the range of 0.92–3.75 ppm are obtained from Landheer et al.[26]. $T_1$ relaxation times at 3T are taken from Hoefemann et al.[27], and water $T_2$ values are averaged over age ranges as reported by Hupfeld et al.[28].

Following the procedure of Wright et al.[25], each MM component is represented as a Voigt lineshape with linewidth parameters derived from the experimental values. The Lorentzian component is calculated from the $T_2$ relaxation times, while the Gaussian component is derived using the measured full-width-at-half-maximum (FWHM) and the following Voigt approximation[29]:

$$f_V \approx 0.5343 f_L + \sqrt{0.2169 f_L^2 + f_G^2} \tag{5}$$

where $f_V$, $f_L$, and $f_G$ are the FWHM values for the Voigt, Lorentzian, and Gaussian components respectively. In this implementation, the Gaussian linewidths are computed without including the shim-related $B_0$ inhomogeneity broadening, which is discussed separately in Section 2.3.6. According to Landheer et al.[26], some $T_2$ values of the MM components could not be reliably quantified. As a result, the corresponding Gaussian linewidths were clipped to a minimum of 1 Hz to ensure numerical stability in the simulations. Relative MM peak amplitudes were also based on experimental spectra reported in the same study.
The resulting base MM spectrum is generated per tissue type and subsequently modulated by a sequence-specific attenuation factor and a user-defined scaling parameter. The total MM signal contribution is given by:

$$S_{\mathrm{mm}}(\nu) = \left\langle \beta_{\mathrm{mm}} \cdot \sum_{\mathrm{l}} L_{\mathrm{l}}(x, y, z) \cdot S_{\mathrm{mm,l}}(\nu) \right\rangle_{x,y,z} \tag{6}$$

where $\beta_{mm}$ is the global MM scaling factor, $L_l(x, y, z)$ is the one-hot encoded tissue label map, and $S_{mm,\ell}(\nu)$ is the MM spectrum for tissue $l$.

### 2.3.3 Residual Water Simulation

The simulation procedure for the residual water signal is based on earlier work that simulated water signals for testing a water removal algorithm[30]. The water signal is represented as a weighted sum of five damped complex exponentials, each with its own amplitude $A_w$, resonance frequency $f_w$, damping constant $\Gamma_w$ (linewidth), and phase $\phi_w$. These components are centered around the typical water resonance range (4.5–4.9 ppm), and their amplitudes and frequencies are adopted from the referenced prior work[30]. The damping constants and phase shifts are randomly drawn from uniform distributions within user-defined ranges, introducing variation across simulations. A separate water-specific amplification factor, $\beta_{water}$, is applied to modulate the relative intensity of the water signal in the simulation.

To reflect realistic lineshapes, the water signal is synthesized twice: once using Lorentzian damping and once using Gaussian damping. The two resulting FIDs are then averaged to form a hybrid signal:

$$FID_{H2O}(t) = \frac{1}{2}\beta_{water} \sum_{w=1}^{5} A_w e^{-\pi \Gamma_w t} e^{i(2\pi f_w t + \phi_w)} + \frac{1}{2}\beta_{water} \sum_{w=1}^{5} A_w e^{-\frac{(\pi \Gamma_w)^2}{4\ln(2)} t^2} e^{i(2\pi f_w t + \phi_w)} \tag{7}$$

The hybrid time-domain signal is Fourier transformed to yield the frequency-domain water spectrum for each tissue label $\ell$:

$$S_{H2O,l}(\nu) = c_\ell \cdot \mathcal{F}[FID_{H2O}(t)] \tag{8}$$

where $c_\ell$ denotes the water concentration for tissue type $\ell$, which is based on literature[31]. This spectrum is spatially weighted by the one-hot encoded label map $L_\ell(x, y, z)$ and averaged over the selected VOI:

$$S_{H2O}(\nu) = \left\langle \sum_{\ell} L_\ell(x, y, z) \cdot S_{H2O,\ell}(\nu) \right\rangle_{x,y,z}. \tag{9}$$

### 2.3.4 Lipid Simulation

When a VOI is selected near the skull, the MRS phantom can simulate lipid contamination in the MRS spectrum. To achieve this, spin systems for biological triglycerides are generated using the SimnTG package[32]. Based on the assumption that human white adipose tissue predominantly consists of oleic acid, palmitic acid, linoleic acid, and palmitoleic acid (together comprising >90% of the total fatty acid content)[33], triglyceride spin systems for these four fatty acids are created, and corresponding spectra are simulated using the FID-A toolbox. The result is a four-component “lipid basis”, $FID_l(t)$, which is used to model lipid contamination. Since lipid signals in MRS spectra tend to exhibit unpredictable phase behavior and broad linewidths, random variations in line broadening $\Gamma_l$, and phase shifts $\phi_l$ are introduced, sampled uniformly from user-defined ranges. The total lipid signal model is given by:

$$S_{\text{lipids}}(\nu) = \left\langle \beta_{\text{lipids}} \cdot M(x,y,z) \sum_{l=1}^{4} \mathcal{F}\left[\text{FID}_l(t) \cdot e^{-\pi\Gamma_l t} \cdot e^{i\phi_l}\right] \right\rangle_{x,y,z} \tag{10}$$

where $\beta_{\text{lipids}}$ is a lipid-specific amplification factor that modulates the relative intensity of the lipid signal and $M(x,y,z)$ is the lipid mask as defined in section 2.1.

### 2.3.5 Noise

To mimic realistic acquisition conditions, complex Gaussian noise is added to the simulated spectra. The noise is generated as zero-mean, white Gaussian noise in both real and imaginary components, independently sampled across all spatial dimensions and frequency points. The noise standard deviation is scaled by a user-defined noise level $\eta$, normalized to the voxel spacing to ensure consistent signal-to-noise behavior across different spatial resolutions. Specifically, the standard deviation is computed as:

$$\sigma_{\text{noise}} = \frac{\eta}{\Delta x} \cdot 10^3 \tag{11}$$

where $\Delta x$ denotes the resolution of the skeleton in millimeters, and the factor 103 ensures realistic noise amplitudes. After generating the full noise volume, it is spatially averaged across the selected VOI to obtain a representative, spatially homogeneous noise profile:

$$S_{\text{noise}}(\nu) = \left\langle \epsilon_{\text{real}}(x,y,z,\nu) + i \cdot \epsilon_{\text{imag}}(x,y,z,\nu) \right\rangle_{x,y,z} \tag{12}$$

with $\epsilon_{\text{real}}, \epsilon_{\text{imag}} \sim \mathcal{N}(0, \sigma_{\text{noise}}^2)$. Since $\eta$ controls the standard deviation per voxel, the resulting SNR effectively increases with the VOI size when the same $\eta$ is used across simulations.

### 2.3.6 Shim Imperfections

The current MRS phantom does not incorporate inherent $B_0$-inhomogeneities, but it provides an option to simulate additional line broadening caused by such imperfections. This is achieved by generating a shim imperfection map. A spatially correlated 3D Gaussian noise field is first generated using a Gaussian filter, where the standard deviation is scaled according to a user-defined correlation length (in mm) and shim amplitude factor (in Hz). To model increased inhomogeneity near tissue boundaries, a signed distance transform is computed between brain tissue and background voxels. A smooth exponential function of this distance produces a boundary weighting term, scaled by an amplification factor. This modulation is used to scale the noise field, enhancing shim variations near tissue edges. The resulting map is zeroed outside the brain volume to confine imperfections to anatomically relevant regions. This shim imperfection map is subsequently applied to all components of the signal model, introducing additional Gaussian line broadening:

$$\mathrm{FID}'_c(r,t) = \mathrm{FID}_c(r,t) \cdot e^{-i2\pi\Delta\omega_r t} \tag{13}$$

where $FID_c(\boldsymbol{r},t)$ is the free induction decay of signal component $c$ at position $\boldsymbol{r}$ within the selected VOI, $\Delta\omega(\boldsymbol{r})$ the shim imperfection map, and $FID_c{}'(\boldsymbol{r},t)$ the affected free induction decay. Figure 3 summarizes the shim imperfection generation process and illustrates its impact on the simulated metabolite signal.

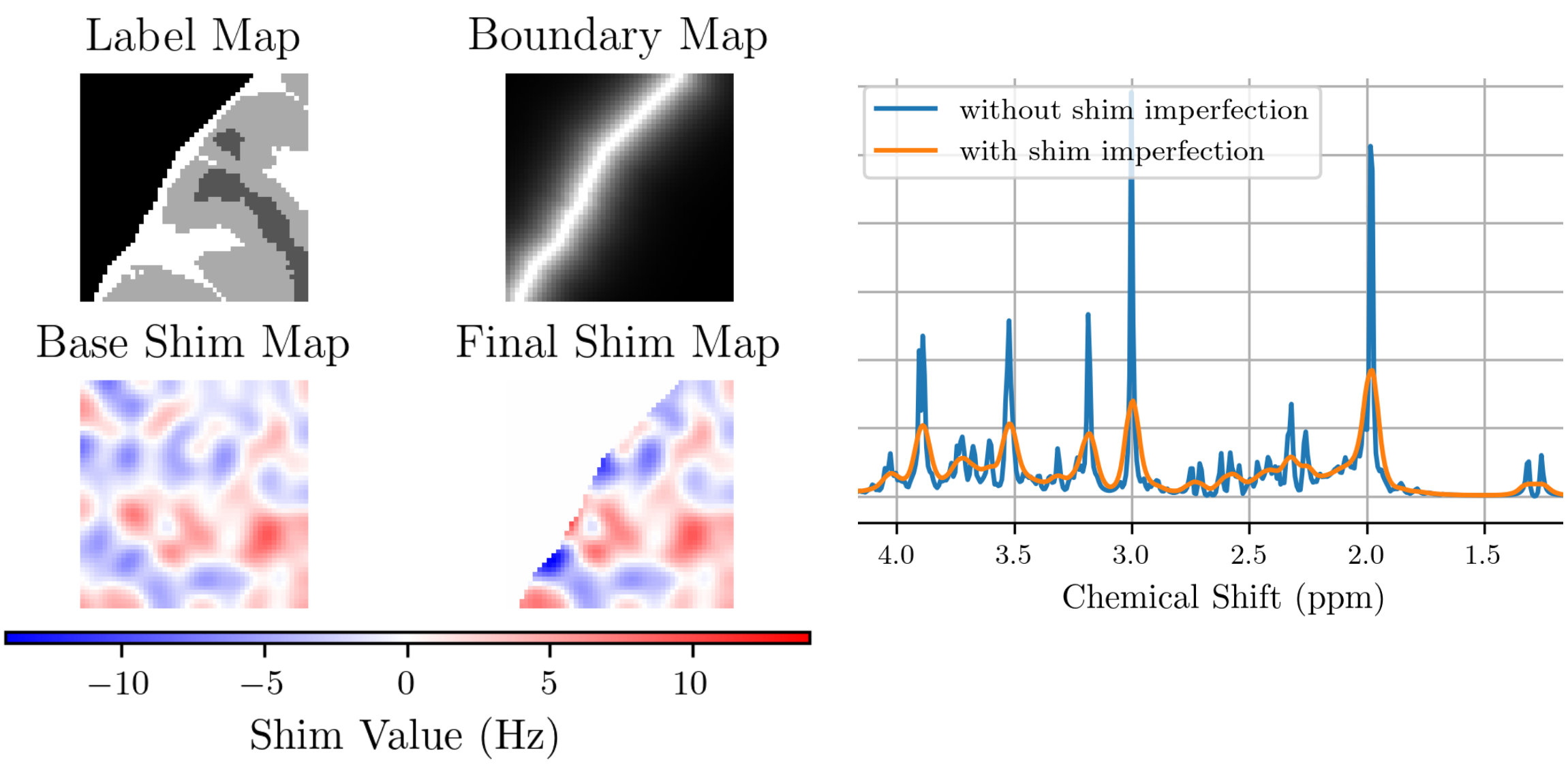

*Figure 3: Procedure for generating a shim imperfection map. The label map displays a 2D slice of the selected VOI with tissue labels. First, a base shim map is created by applying spatially correlated 3D Gaussian noise with a user-defined correlation length. Next, a boundary weighting map enhances shim imperfections near the tissue–air interface to produce the final shim map. The spectrum on the right compares a metabolite signal without shim imperfections (blue) to one with imperfections applied (orange), illustrating the resulting line broadening.*

### 2.3.7 Simulation Outputs

Upon completion of the simulation, the output consists of fully synthesized MRS spectra in which all individual signal components (e.g., metabolites, macromolecules, water, lipids, and noise) are preserved separately. This enables component-wise analysis or recombination, depending on the desired downstream application. In addition to the spectral data, a comprehensive configuration file is generated in JSON format. This file stores all simulation parameters, including those listed in Table 1, ensuring full reproducibility and traceability of each simulation run.

All simulation results can be exported in the NIfTI-MRS data format[34], which is compatible with standard neuroimaging pipelines and facilitates integration with existing software tools. This standardized format supports reproducible software development and facilitates the efficient training and evaluation of machine learning models using simulated ground truth data.

Default parameters for the simulation pipeline are embedded in the codebase and are designed to replicate realistic in-vivo conditions at 3T. These defaults can be overridden via user-supplied configuration files to accommodate different experimental setups or research needs.

## 2.4 Evaluation Methods

To assess the realism and flexibility of the proposed MRS phantom simulation pipeline, a series of evaluations is conducted. These include simulations under varying acquisition conditions and comparisons with in-vivo data.

### 2.4.1 Variation Across Simulation Parameters

To demonstrate the adaptability of the MRS phantom, multiple sets of simulated spectra are generated by systematically varying key acquisition-related parameters. Specifically, simulations are conducted across a range of signal-to-noise ratios (SNRs), linewidths, and levels of lipid contamination and residual water signal. These variations are intended to reflect typical in-vivo conditions as well as more extreme cases that may arise under suboptimal acquisition settings. By covering a broad parameter space, the phantom enables targeted testing of algorithmic robustness and sensitivity to (acquisition-related) artifacts.

### 2.4.2 Comparison with In-Vivo Data

To evaluate the realism and utility of the MRS phantom, a comparative analysis is performed using in-vivo data from a previously published study[35]. Spectra are selected from a dataset acquired on a Philips 3T system (Philips Healthcare, Best, the Netherlands) using the following parameters: TR = 2000 ms, TE = 35 ms, a spectral bandwidth of 2000 Hz, and 2048 spectral points. Metabolite concentrations from these in-vivo spectra are quantified using the Osprey toolbox[6], and the resulting estimates serve as input for the phantom to generate simulated spectra that closely mimic the corresponding in-vivo acquisitions. The simulations replicate the acquisition and sequence parameters of the in-vivo dataset, and the remaining parameters are optimized to maximize similarity between the simulated and in-vivo spectra.

Both simulated and in-vivo spectra are preprocessed using FSL-MRS[7], which includes residual water removal, phase and frequency correction, baseline correction, and normalization. The dynamics of the in-vivo spectra are averaged prior to this preprocessing. To further evaluate representational similarity, both datasets are embedded into a low-dimensional feature space using t-distributed stochastic neighbor embedding (t-SNE)[36]. This dimensionality reduction technique facilitates evaluation of whether the simulated spectra occupy a similar region in feature space as the in-vivo data, indicating representational similarity.

Finally, the simulated spectra are quantified using Osprey, allowing direct comparison of quantification results between simulated and in-vivo datasets. This step helps validate both the spectral realism and the quantitative accuracy of the simulation pipeline.

# 3 Results

## 3.1 MRS Phantom

Filtering of the metabolite database resulted in a metabolite concentration dataframe with 783 entries, consisting of 54 unique references and including 19 unique metabolites. For the $T_2$ relaxation times, 219 entries are available after filtering with 18 references and 19 metabolites. All these entries are used to determine the metabolite information in the MRS phantom. For some metabolites, missing data resulted in incomplete values for gray matter (GM) and/or white matter (WM) concentrations and $T_2$ relaxation times.

Figure 4 displays the developed graphical user interface (GUI) for interacting with the digital MRS phantom. The left panel allows users to select the skeleton, metabolite dataframe, and basis set. The upper right panel enables voxel placement, while the bottom panel provides access to simulation parameters. These parameters are organized into multiple tabs, each corresponding to a specific signal component. The resulting spectrum is displayed within the same window, with options to visualize individual signal components and to export the data in the NIfTI-MRS format.

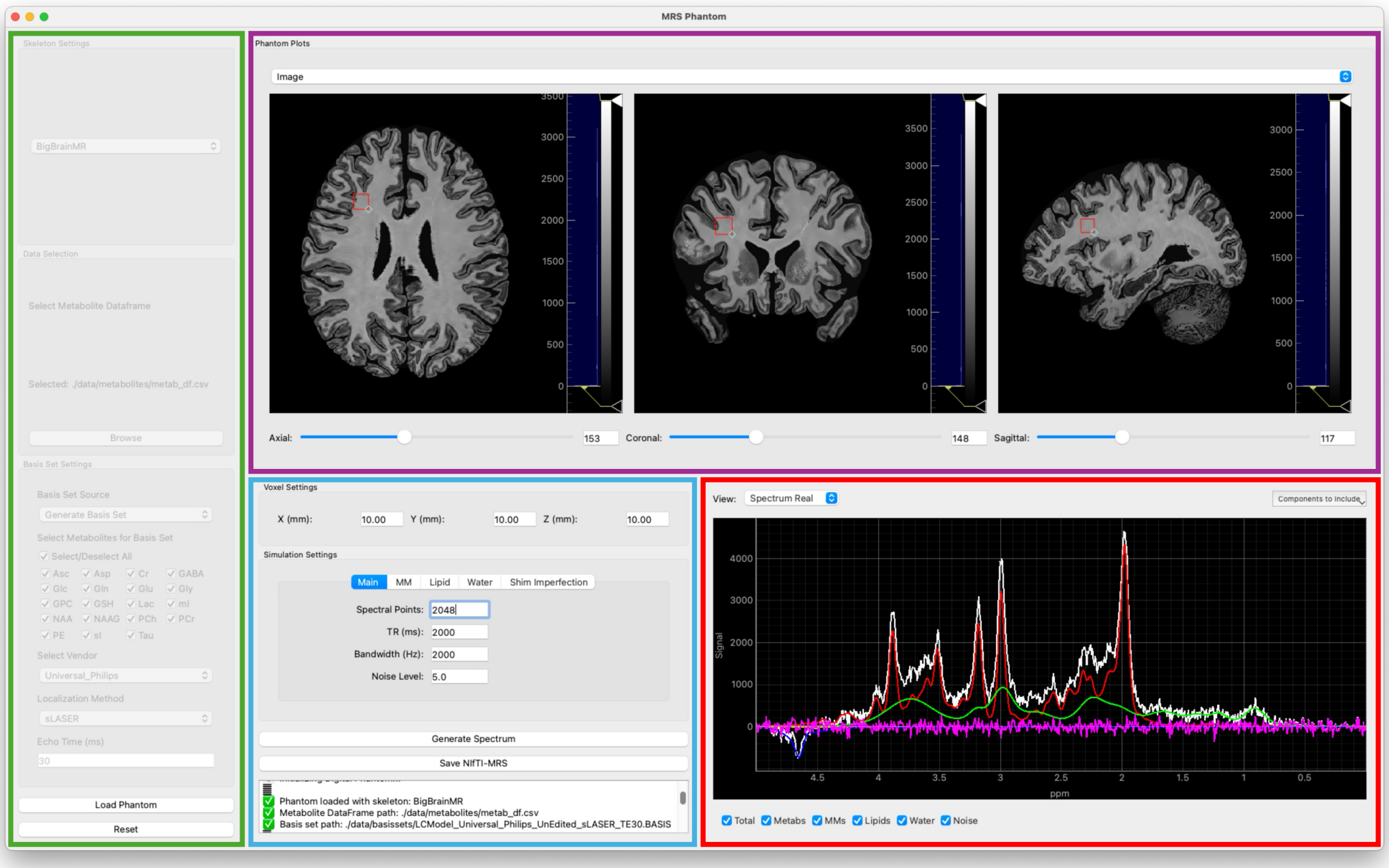

*Figure 4: Screenshot of the graphical user interface (GUI) of the digital MRS phantom. The left panel (green box) displays settings for the skeleton, metabolite dataframe, and basis set. The top panel (purple box) shows the three orthogonal brain views used for voxel placement. The bottom-middle section (blue box) contains the simulation settings panel and message box. The right panel (red box) visualizes the simulated spectrum, including its individual signal components.*

## 3.2 Variation Across Simulation Parameters

The MRS phantom is used to generate a range of simulated spectra by varying key acquisition-related parameters, including noise levels, lipid contamination, residual water signal, and shim imperfection settings. Figure 5 presents a selection of spectra simulated under different parameter configurations, with annotations indicating the corresponding values used for each.

All spectra are generated using the BigBrainMR skeleton, the metabolite dataframe described in Section 2.2.1, and a basis set for sLASER localization at an echo time (TE) of 30 ms, based on universal Philips pulse sequences. An example configuration file for simulating lipid contamination variations is provided in the Supporting Information (Figure S2), while all other configuration files are available in the GitHub repository.

In each case, the figure displays the value of the parameter that is varied. For lipid contamination and shim imperfections, two parameters are manipulated. Lipid contamination is controlled by an amplitude factor and the spatial smoothing width ('lipid_amp_factor' and 'lipid_sigma'; see Table 1). Shim imperfections are simulated by generating a spatially correlated distortion field, where the smoothness is governed by a correlation length and the distortion amplitude is scaled by a shim amplitude parameter ('shim_corr_length' and 'shim_amplitude_hz'; see Table 1).

The simulated spectra show clear and interpretable effects across these variations. Increasing the noise level results in spectra with lower SNR. Greater lipid contamination, either due to higher amplitude or broader smoothing, leads to more pronounced lipid artifacts. Shim imperfections with larger amplitude values result in broader linewidths, while increasing the correlation length produces more subtle effects, though larger values can lead to noticeable frequency shifts. Similarly, increasing the residual water amplitude leads to a visibly larger water peak.

It is important to note that the simulation of both residual water and lipid signals includes stochastic components. As a result, spectra generated with identical parameter settings may still exhibit minor variations.

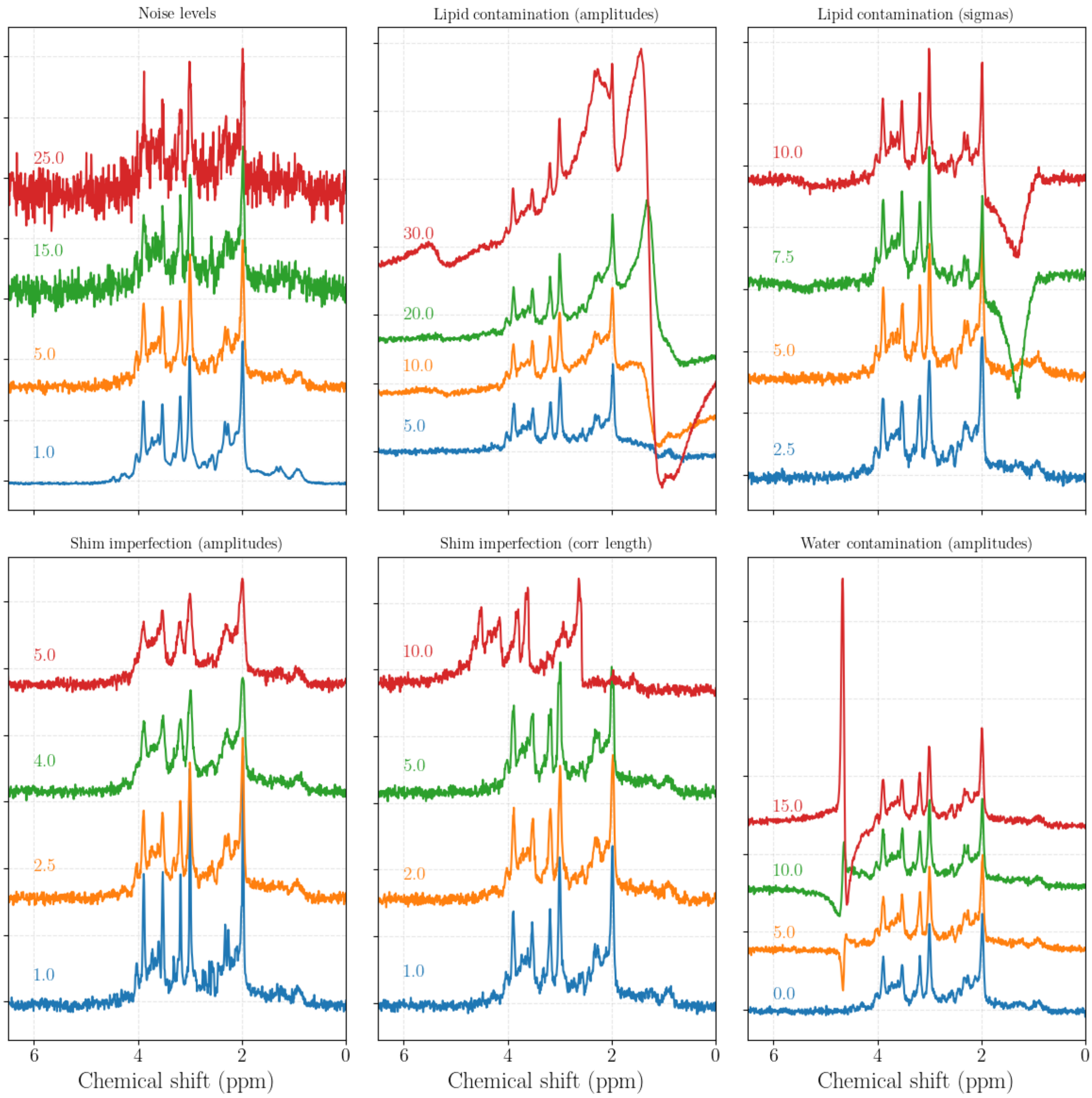

*Figure 5: Simulated spectra illustrating the effect of varying key simulation parameters. The top row shows variations in noise levels, lipid amplitude factors (lipid_amp_factor), and the lipid mask smoothing parameters (lipid_sigma). The bottom row displays spectra with different shim imperfection amplitudes (shim_amplitude_hz) and correlation lengths (shim_corr_length), as well as variations in the amplitude factor for residual water signals. All plots demonstrate gradual and interpretable changes in their respective signal components.*

## 3.3 Comparison with In-Vivo Data

A total of 104 in-vivo spectra and 480 simulated spectra were analyzed, with the latter generated under similar acquisition conditions. Figure 6 compares their mean and standard deviation. Simulated spectra closely resemble in-vivo spectra in shape and variability within the metabolite-dominated 2–4 ppm range, though some simulated peaks show slightly higher variability. Larger discrepancies appear outside this window: in the 0–2 ppm region, associated with lipid signals, simulated spectra exhibit greater variability than in-vivo data. Conversely, in the 4–6 ppm range, where residual water signals often persist, in-vivo spectra show more variation despite preprocessing applied to both datasets to reduce water contamination. These results highlight distinct patterns of variability related to lipids and residual water in the two datasets.

Figure 7 shows a t-SNE embedding of in-vivo and simulated spectra based on PCA-extracted features. Several simulated spectra cluster with in-vivo data, indicating that the simulation captures key spectral characteristics. For example, Figure 7a shows a simulated and in-vivo pair that are close in feature space and visually similar. However, some in-vivo spectra form distinct clusters not overlapped by simulated data, often due to prominent residual water peaks, as shown in Figure 7b.
Simulated spectra cover a broader region of the t-SNE space, reflecting higher variability across the parameter space. Figure 7c shows examples with diverse lipid profiles, consistent with the lipid-related variability seen in Figure 6.

Figure 8 shows quantification results for both in-vivo and simulated spectra. Metabolite amplitudes from simulated data generally overlap with in-vivo values, suggesting that the simulation captures realistic concentration ranges. Despite this agreement, Mann–Whitney U tests indicate significant differences ($p < 0.05$, marked with *) for several metabolites, macromolecules, and lipid components. These differences likely stem from the larger simulated sample size (480 vs. 104 spectra) and greater spectral variability, as seen in Figure 7. Quality metrics for the creatine peak show comparable results: in-vivo spectra have an SNR of 143 ± 39 and FWHM of 6.34 ± 1.21 Hz, while simulated spectra yield an SNR of 107 ± 11 and FWHM of 6.32 ± 0.63 Hz.

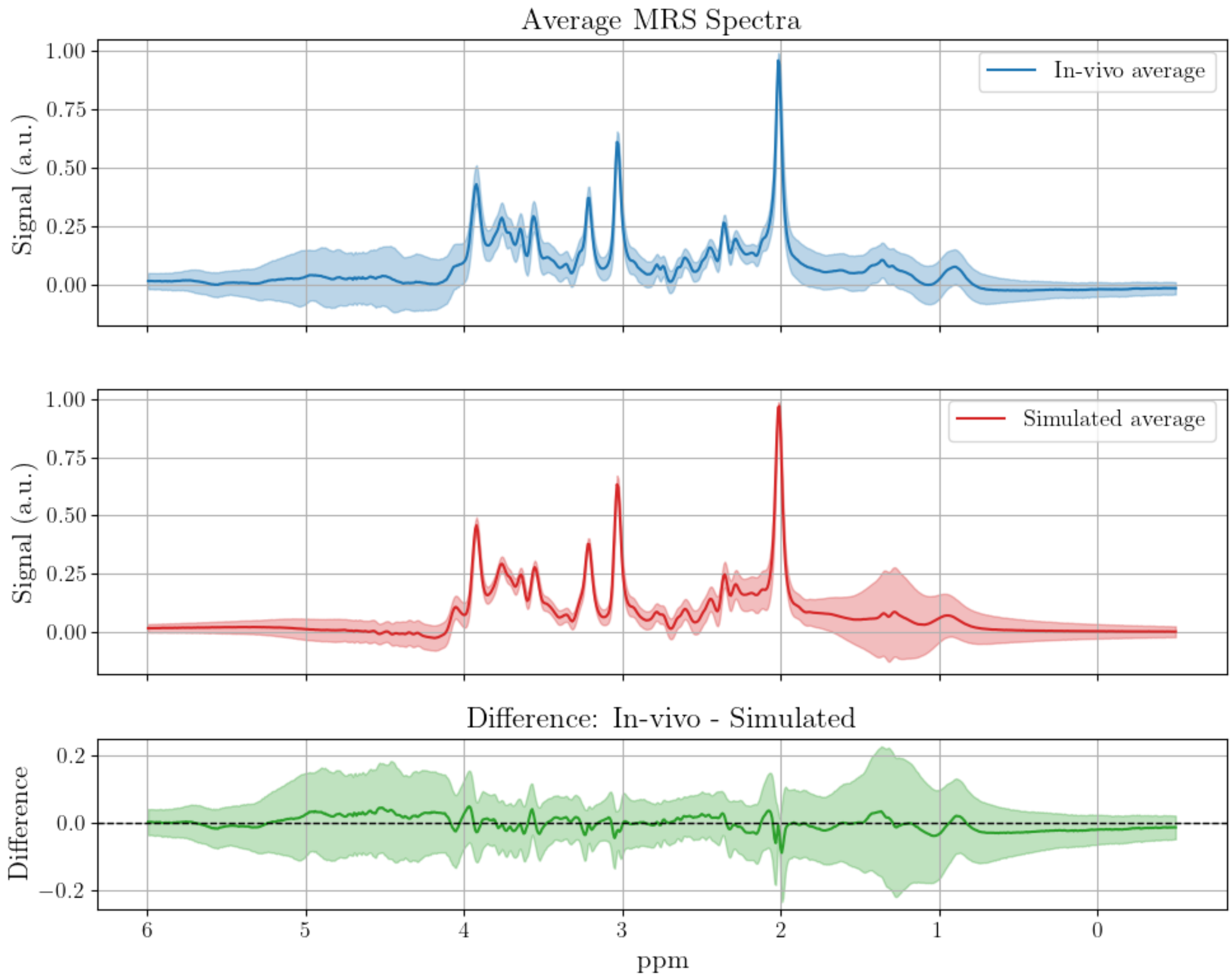

*Figure 6: Averaged spectra for the in-vivo (top, blue) and simulated (middle, red) datasets with their corresponding differences (bottom, green). The shaded areas indicate the standard deviation of the signal.*

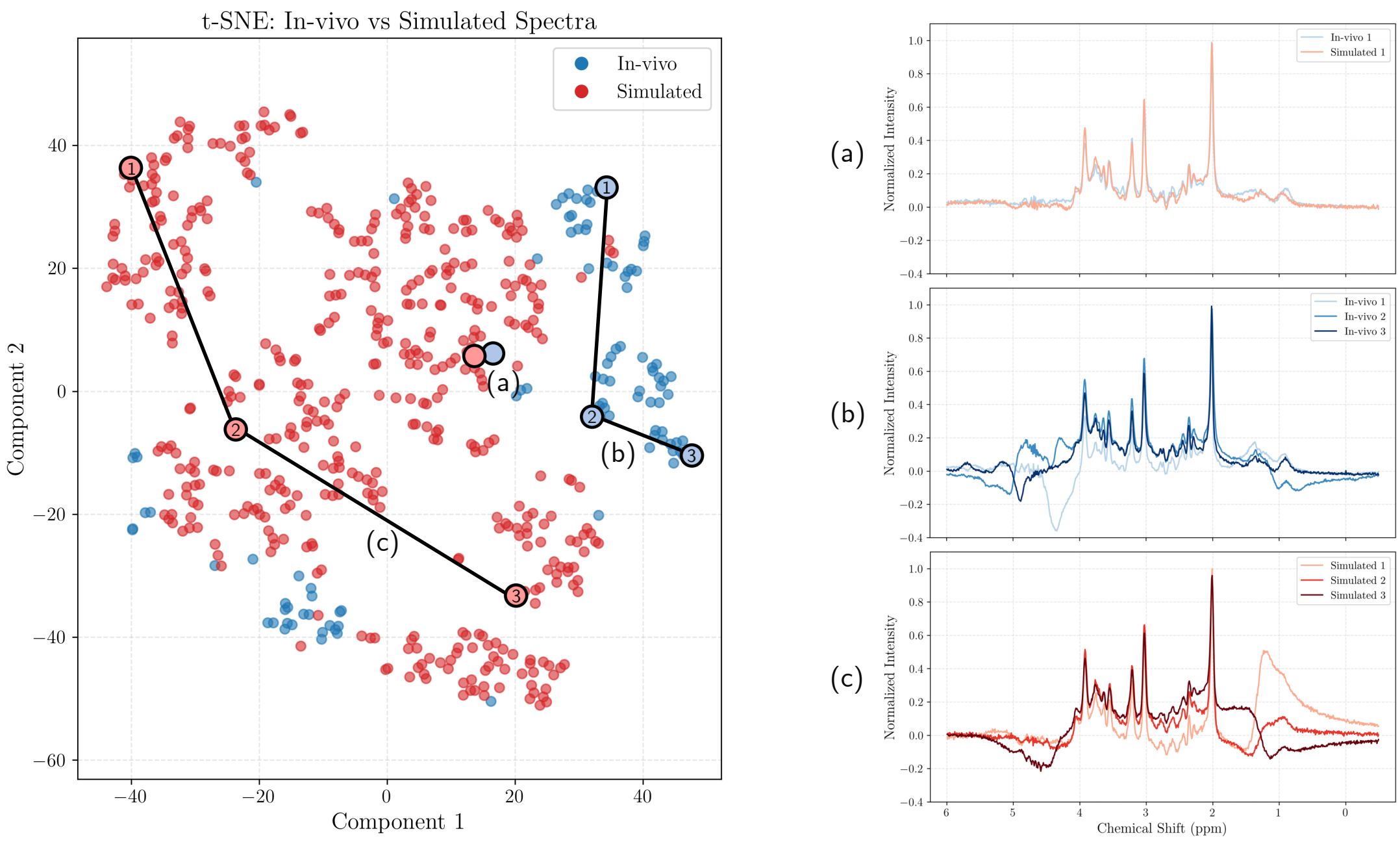

*Figure 7: t-SNE plot showing the feature space of in-vivo and simulated spectra, alongside representative example spectra.* ***(a)*** *Overlapping points illustrate simulated spectra that closely resemble in-vivo spectra.* ***(b)*** *Distinct in-vivo clusters indicate spectral features not fully captured by the simulations, often characterized by varying residual water*

*signals. **(c)** Simulated spectra span a broader range in feature space, with notable variation in lipid contamination levels.*

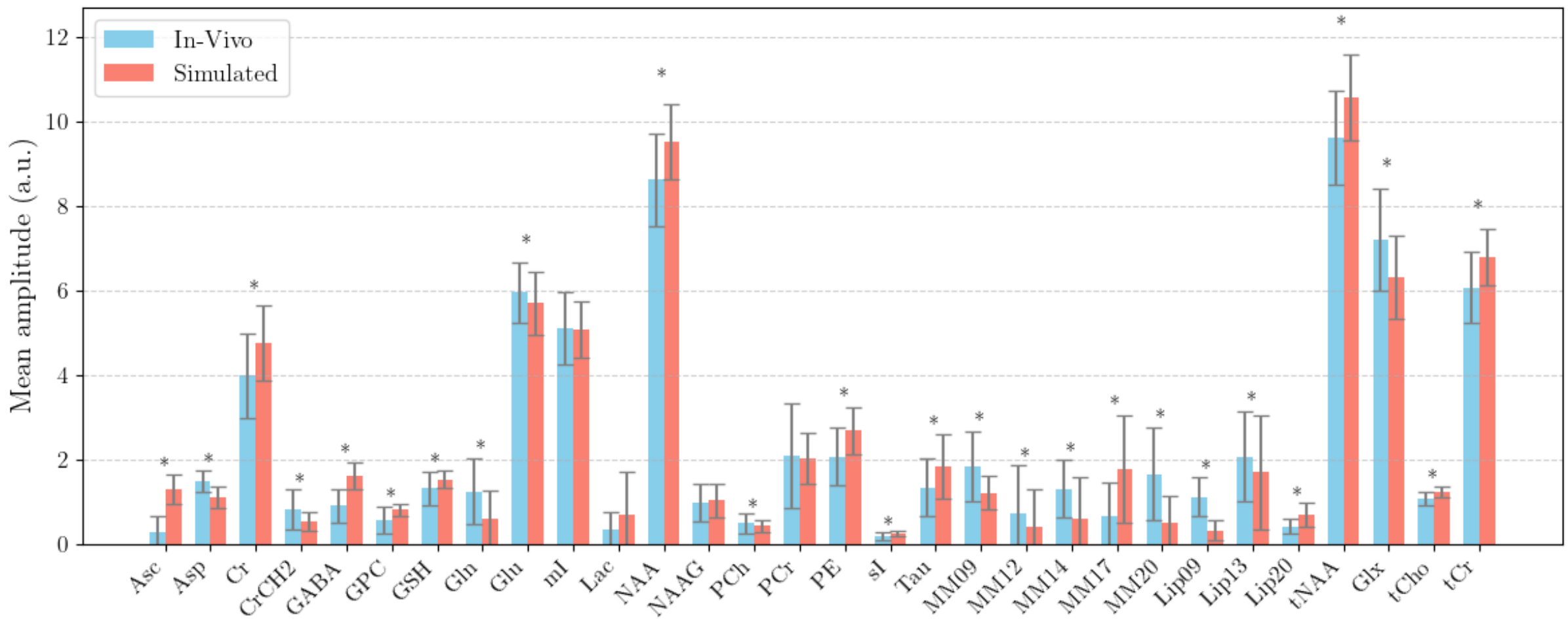

*Figure 8: Quantification results for in-vivo and simulated spectra, reported as mean metabolite amplitudes (arbitrary units). Error bars indicate standard deviations across subjects. Statistical significance of group differences is shown above each bar ($p < 0.05$ indicated with *).*

# 4 Discussion

The proposed MRS digital phantom framework is a modular tool for simulating realistic MRS data using tissue-specific metabolite information and a signal-based model. All signal components are based on physics-informed models or prior knowledge. Exporting data in NIfTI-MRS format ensures broad compatibility, making the framework suitable for augmenting in-vivo datasets or generating synthetic data for algorithm development and robustness testing.

## 4.1 MRS Phantom

The framework is tested with two anatomical templates (BigBrain-MR and MRiLab), providing a solid foundation for benchmarking. Its modular design facilitates future expansion to diverse anatomical models, enhancing versatility. Efficient management of multiple templates and integrating a registration method to align anatomical skeletons would improve dataset consistency. For instance, registering skull or lipid voxels between templates could standardize key features such as lipid map creation. Adding further maps, like $B_0$-inhomogeneity, would provide valuable prior knowledge for simulations. Currently, shim imperfections are modeled with a basic simulation approach. Future versions could replace this with detailed inhomogeneity maps.

The metabolite dataframe used is derived from a subset of a larger meta-analysis database. As shown in Figure 2, this approach balances the need for data homogeneity with inclusion of enough studies to yield meaningful averages. Although summarizing heterogeneous sources poses challenges, the framework remains flexible, allowing adjustment of metabolite means and standard deviations to simulate pathological conditions. Incorporating metabolite data from diverse patient cohorts and tissue types would further enhance the framework’s applicability and clinical relevance.

Currently, the framework does not account for spatial variability in metabolite concentrations within brain regions, despite evidence indicating such heterogeneity exists[23,37,38]. Including such variability would enhance the utility of the framework for investigating localized metabolic differences and disease-specific alterations. While the current metabolite dataframe already includes variations for GM, WM and CSF, future improvements could extend this dataframe for specific brain regions when more literature is available on metabolite concentration and relaxation time differences per brain region.

## 4.2 Signal Model and In-Vivo Comparison

The signal model implemented in the MRS phantom provides a robust foundation for generating realistic MRS data, integrating seamlessly with the broader simulation framework. While it already supports a variety of use cases, several aspects can be improved to more accurately reflect real-world acquisition conditions.

The MRS phantom is designed to accept any pre-generated basis set and includes built-in integration with MRSCloud for on-the-fly basis set generation if none is provided. Although basis set generation is not the primary focus of this work, this flexibility demonstrates compatibility with diverse workflows and acquisition protocols.

Currently, the macromolecule (MM) signals are simulated based on literature data obtained at 3T. While the model allows for different relaxation parameters for GM and WM, existing literature does not yet report tissue-specific MM differences at 3T. As such, the MM model remains a general approximation in terms of tissue-specific spectra and could be refined as new published data becomes available.

The residual water and lipid components are simulated using random line-broadening and phase shifts to capture their stochastic nature. This approach is effective in producing variability (Figure 5), but when aiming to replicate specific in-vivo conditions, this randomness may fall short (Figure 6). These components often follow structured patterns in real data. Future improvements could include enhanced physics-based modeling or data-driven parameter sampling, leveraging in-vivo examples to guide the realism of these contaminations.

Figure 7 demonstrates the broad spectral variability produced by the phantom, ranging from in-vivo-like spectra to more extreme cases. This diversity enables applications such as algorithm robustness testing and data augmentation. However, as Figure 8 shows, statistically significant differences in metabolite quantification remain between simulated and in-vivo data. Although these differences are statistically significant, they may not be practically meaningful. Simulated metabolite concentrations are sampled around the in-vivo mean, but variability in lipid and water contamination, particularly when not well matched to in-vivo conditions, can affect quantification accuracy. Figure 7 illustrates this, with some simulated spectra closely resembling in-vivo data while others deviate substantially.

Additionally, the simulation currently models ten MM components, while Osprey's default quantification pipeline fits five. This mismatch, along with empirical tuning of MM amplitudes in the simulation, may further contribute to differences in quantification. A more detailed and systematic optimization of the MM signal model could improve similarity and quantification accuracy.

To further improve the realism and versatility of the MRS phantom, future extensions could include the simulation of acquisition-related artifacts. For example, eddy currents, which introduce phase shifts and baseline distortions, are a common artifact that can affect spectral quality and quantification accuracy. Incorporating such effects would enable users to test and validate correction algorithms in a controlled environment.

In addition, this framework could be extended to support MRSI simulations, further increasing its applicability. This would allow for the modeling of spatial artifacts and heterogeneity across voxels, making the phantom suitable for testing multi-voxel analysis pipelines and developing spatial correction techniques.

### 4.3 Computation and Accessibility

All simulations in this work are performed on standard local hardware, demonstrating that the MRS phantom does not require high-performance computing resources. The developed GUI further enhances accessibility by allowing users to easily configure simulations and visually interpret the results. For large-scale data generation, the phantom can also be run without the GUI using configuration files, enabling more efficient batch processing. All code is openly available to promote transparency, reproducibility, and further development by the MRS research community.

## 5 Conclusion

This work presents a novel and modular framework for simulating MRS data using a digital brain phantom. By integrating anatomical priors and metabolite distributions with a flexible, signal-based simulation model, the framework enables the generation of realistic, tissue-specific spectra. Comparisons with in-vivo data demonstrate its ability not only to mimic physiological MRS signals but also to produce broader spectral variability, including extreme cases valuable for testing the robustness of quantification algorithms and machine learning models. The phantom's compatibility with open formats and its accessibility through both a GUI and script-based workflows make it a practical tool for a wide range of users. Its open-source nature and modular architecture encourage future extensions, allowing the framework to evolve alongside the needs of the MRS research community.

## 6 Acknowledgements

We would like to thank the Fulbright Commission the Netherlands for financially supporting the visit of D.M.J. van de Sande to The Johns Hopkins University School of Medicine, Baltimore, USA. This research was performed in the Spectralligence project, funded by the European ITEA4 program (project 20209).

# Supporting Information

Additional supporting information may be found in the online version of the article at the publisher's website

Table S1: Example of the metabolite dataframe that is integrated in the MRS phantom.
Figure S2: Example configuration file used to generate simulated spectra illustrating lipid contamination variations (see Figure 5).

# Supporting Information

Supporting Information Table S1: Example of the metabolite dataframe that is integrated in the MRS phantom.

Supporting Information Figure S2: Configuration file used for generating various spectra.

***Table S1:*** *Example of the metabolite dataframe for two metabolites: mI and NAA. For each metabolite there are as many entries as labels that are used in the MRS Phantom. All values for WM and GM are based on the used literature study, CSF values are manually set, and all background values are set to 0.0. $T_1$ values are not added yet but have been implemented as placeholder for future updates.*

| Metabolite | Label | Tissue | Conc_mean [mM/IU] | Conc_std [mM/IU] | $T_1$ [ms] | $T_2$ [ms] |
|---|---|---|---|---|---|---|
| **mI** | 0 | Background | 0.0 | 0.0 | 0.0 | 0.0 |
| **mI** | 1 | WM | 5.42 | 0.63 | - | 189.90 |
| **mI** | 2 | GM | 4.86 | 0.27 | - | 200.03 |
| **mI** | 3 | CSF | - | - | - | - |
| **NAA** | 0 | Background | 0.0 | 0.0 | 0.0 | 0.0 |
| **NAA** | 1 | WM | 8.76 | 1.1 | - | 291.44 |
| **NAA** | 2 | GM | 8.33 | 0.59 | - | 265.31 |
| **NAA** | 3 | CSF | - | - | - | - |

***Figure S2:*** *Example configuration file used to generate simulated spectra illustrating lipid contamination variations (see Figure 5). Additional configuration files for other simulation parameters are available in the GitHub repository.*

```
{
    "skeleton": "BigBrainMR",
    "path2metabs": "data/metabolites/metab_df.csv",
    "basis_set_dir": "data/basissets/",
    "path2basis":
"data/basissets/LCModel_Universal_Philips_UnEdited_sLASER_TE30.BASIS",
    "metabs": [
        [],[],[],[],[]
    ],

    "basis_set_settings":{
        "vendor": "Universal_Philips",
        "localization": "sLASER",
        "TE": 30
    },

    "simulation_params": {
        "spectral_points": 2048,
        "TR": 2000.0,
        "bandwidth": 2000.0,
```

```
        "noise_level": 5.0,

        "mm_level": 25.0,
        "mm_json_file": "./data/macromolecules/mm_params.json",

        "lipid_amp_factor": [5.0, 10.0 , 20.0, 30.0],
        "lipid_sigma": [2.5, 5.0, 7.5, 10.0],
        "lipid_phase_min": -180.0,
        "lipid_phase_max": 180.0,
        "lipid_lw_min": 30.0,
        "lipid_lw_max": 50.0,

        "water_amp_factor": 0.0,
        "water_phase_min": -180.0,
        "water_phase_max": 180.0,
        "water_damping_min": 0.0,
        "water_damping_max": 50.0,

        "shim_amplitude_hz": 2.5,
        "shim_corr_length": 1.0,
        "shim_boundary_amp_factor": 1.5,
        "shim_boundary_smoothing": 2.0
    },
        "voxel_definitions": [
        {
            "coords": [
                110,
                135,
                402,
                427,
                191,
                216
            ],
            "size": [
                25,
                25,
                25
            ],
            "size_mm": [
                10.000000149011612,
                10.000000149011612,
                10.000000149011612
            ]
        }
    ]

}
```